\documentclass[12pt,epsfig,amsmath]{article}
\usepackage{graphicx,epsfig}
\newcommand{\slashed}[1]{~\slash \mbox{$\!\!\!$}{#1}}

\begin{document}

%\begin{titlepage}
%\begin{flushright}
%PUPT-2010    \\
%hep-ph/0601
%\end{flushright}

%\vspace{7 mm}

\vspace{7 mm}

\title{Single spin asymmetry and five-quark components of the proton}
\author{
F.~X.~Wei~$^{1,2)}$\thanks{weifx@ihep.ac.cn},
B.~S.~Zou$^{1,2,3)}$\thanks{zoubs@ihep.ac.cn}  \\
1) Institute of High Energy Physics and Theoretical Physics Center \\ for Science Facilities, CAS, Beijing 100049\\
2) Graduate University, Chinese Academy of Sciences, Beijing
100049 \\
3) Center of Theoretical Nuclear Physics, National Laboratory of\\
Heavy Ion Accelerator, Lanzhou 730000 }
\date{July 9, 2008}
\maketitle

\begin{abstract}
We examine the single-spin asymmetry (SSA) caused by the five-quark
components of the proton for semi-inclusive electroproduction of
charged pions in deep-inelastic scattering on a transversely
polarized hydrogen target. The large SSA is considered to have close
relation with quark orbital motion in the proton and suggests that
the quark orbital angular momentum is nonzero. For the five-quark
$qqqq\bar{q}$ components of the proton, the lowest configurations
with $qqqq$ system orbitally excited and the $\bar{q}$ in the ground
state would give spin-orbit correlations naturally for the quarks in
a polarized proton. We show that based on the basic reaction $\gamma
q \to \pi q'$, the orbital-spin coupling of the probed quarks in the
five-quark configuration leads to the single-spin asymmetry
consistent with recent experiment results.
\end{abstract}

%\end{titlepage}
\section{Introduction}

The spin composition of the proton in terms of its fundamental quark
and gluon degrees of freedom is a central focus of proton structure.
Whether the quark orbital angular momentum is zero or not is one of
the key points to solve this problem. The importance of quark
orbital angular momentum, which one might have taken to vanish in
the ground state, has been evident since the work of
Sehgal~\cite{seh74}. The orbital angular momentum structure of the
proton is of considerable interest and much effort has gone into
devising ways to measure it. Recently, the investigation of
single-spin asymmetries in hadronic processes suggests that the
orbital angular momentum of quark in the proton is nonzero. This
puts forward a challenging opportunity and imposes an important
constraint on phenomenological studies.

The measurements from the HERMES and SMC Collaborations show a
remarkably large single-spin asymmetry of the proton in
semi-inclusive deep inelastic lepton scattering $lp^{\uparrow} \to
l'\pi X$~\cite{air05,air00}. Large single-spin asymmetries have also
been seen in hadronic reactions such as $p\bar{p}^{\uparrow}\to\pi
X$~\cite{bra96}, where the antiproton is polarized normal to the
pion production plane, and $pp\to \Lambda^{\uparrow}X$~\cite{hel97},
where the hyperon is polarized normal to the production plane. We
focus on the semi-inclusive pion leptoproduction $lp^{\uparrow}
\rightarrow l'\pi X$, because of the simplification with only one
baryon in the initial state. To describe the single-spin asymmetries
for the semi-inclusive deep inelastic lepton scattering
$lp^{\uparrow} \to l \pi X$, a non-zero Sivers function
$f_{1T}^{\bot}$  is usually used to reflect the asymmetric
transverse momentum distribution of the quark in a transversely
polarized proton. The Sivers function $f_{1T}^{\bot}$ is
proportional to $f^{\uparrow} - f^{\downarrow}$, i.e.,
$f_{1T}^{\bot} \sim f^{\uparrow} - f^{\downarrow}$, where
$f^{\uparrow}$ and $f^{\downarrow}$ is the quark distributions in
the transversely polarized proton for the direction up and down,
respectively. The non-zero Sivers function means that unpolarized
quark distribution is different in up and down nucleon spin states.
This difference can only originate from the coupling of quark spin
and its orbital motion, and suggests non-zero quark orbital angular
momentum which can contribute to spin structure of the nucleon.

Based on the nonzero quark orbital angular momentum and final or
internal interaction, some theoretical predictions are available in
the framework of MIT bag model~\cite{yua03} and quark-diquark
configurations~\cite{bro02,bac04,zhu04}. The Sivers functions
obtained in these two models are different in sign and magnitude. It
is suggested that the difference could be originated from sea-quark
contributions~\cite{bac04}, which have opposite sign to that from
valence quarks. Recently, the five-quark components of specific
configurations have been proposed and carried out in various aspects
to understand the spin-flavor structure of proton and other
baryons~\cite{zou05,an206,ris06,an06}. For the proton, the possible
five-quark configurations consist of $uudu\bar{u}$, $uudd\bar{d}$,
$uuds\bar{s}$, and so on. In these configurations, the ones with
$\bar{q}$ ($q=u,d,s$) in its ground state and the $uudq$ subsystem
in P-state give proper results. Then one can find that each $u$ or
$d$ have $\frac{1}{4}$ probability to be in P-state. This gives a
natural explanation of non-zero quark orbital angular momentum in
the nucleon. In this paper, we examine the single-spin asymmetry
caused by the five-quark components in the proton.

\section{The orbital angular momentum of quarks and the $qqqq\bar q$ components in the proton}

The importance of the quark orbital angular momentum in nucleon
structure has been discussed by Sehgal~\cite{seh74} and
Ratcliffe~\cite{rat87} in different contexts. The proton's ``spin
crisis"~\cite{ash88} indicates the presence of quark orbital angular
momentum. The probe of quark orbital momentum has become available
after the detailed investigation of the composition of nucleon spin
in terms of quark and gluon~\cite{jaf90} and the generalized parton
distributions (GPDs)~\cite{ji98}. The SSA can be viewed as one of
the observable to measure the effect of quark angular momentum
through the spin-orbit correlations~\cite{bur07}. There have been
two mechanisms proposed beyond the naive parton model: the twist-3
quark-gluon-quark correlations and transverse momentum dependent
(TMD) parton distributions, which is related to the quark orbital
angular momentum in nucleon. The intrinsic transverse momentum of
quarks is crucial to generate large SSA, because the nonzero quark
orbital angular momentum can lead to the flip of the hadron
helicity. There have already been many works about the relations
between quark orbital motion and SSAs in different
processes~\cite{yua03,bro02,bor93,lia00,ji03} with the conclusion
that the nonzero quark orbital angular momentum is necessary to
produce the large SSA.

However, in the classic qqq configuration of the nucleon, all quarks
are in the S-wave and give a zero orbital angular momentum. Hence
one must go beyond the simple qqq quark model by including
additional $q\bar{q}$ pair(s). The five-quark $uudq\bar{q}$
components in the proton with the $\bar q$ of negative intrinsic
parity demands either a quark or anti-quark in the P-wave to make up
the proton of positive intrinsic parity. So it gives naturally the
nonzero quark orbital angular momentum.

In the five-quark component model~\cite{zou05,an206,ris06,an06},
there are significant $uudd\bar d$ and $uuds\bar s$ components in
the proton with the anti-quark in the orbital ground state and the
four quarks in the mixed orbital $[31]_X$ symmetry, i.e., one in
P-wave and three in S-wave, together with flavor-spin
$[4]_{FS}[22]_F[22]_S$ symmetry. This kind of configuration is found
to have the lowest energy no matter whether the hyperfine
interaction between quarks is described by the color magnetic
interaction or by the flavor and spin dependent hyperfine
interaction of chiral quark model \cite{Helminen}. In fact, this
configuration is very similar to the Jaffe-Wilczek's diquark
configuration for penta-quarks \cite{Jaffe}. Only two quarks with
different flavors can form a good diquark. Obviously the $uudu\bar
u$ cannot form this kind of configuration and can only in
configurations with higher energies. Hence in the proton, there is
less $uudu\bar u$ component than $uudd\bar d$ component. Therefore,
the quark wave function for the proton may then be expanded as:
\begin{equation}
|p> = A_{3q}|uud> + A_{d\bar d} |[ud][ud]\bar d> + A_{s\bar s}
|[ud][us]\bar s> + A_{u\bar u}|uudu\bar u>\,
\end{equation}
with the normalization condition $|A_{3q}|^2+|A_{d\bar d}|^2+
|A^2_{s\bar s}|^2+|A^2_{u\bar u}|^2=1$. Define $P_{q\bar q}\equiv
|A_{q\bar q}|^2$, which represents the probability to find the
$uudq\bar{q}$ component in a proton with $q=u$, $d$ or $s$. Then to
reproduce the observed \cite{Garvey} light flavor sea quark
asymmetry in the proton, $\bar d-\bar u=0.12$, one has $P_{d\bar
d}-P_{u\bar u} =12\%$. To reproduce the observed \cite{Florian}
strangeness spin of the proton, $\Delta_s=-0.10\pm 0.06$, one needs
$P_{s\bar s} = (12-48)\%$ \cite{an206}.

There is another constraint on the percentage of strange quarks in
the proton, coming from a next-to-leading-order QCD analysis of
neutrino charm production \cite{baz95}. It was found that the
nucleon strange quark content is suppressed with respect to the
non-strange sea quarks by a factor $\kappa = 0.477 \:
^{+\:0.063}_{-\:0.053}$ under the assumption of $s$-$\bar s$
symmetry. If allowing $s$-$\bar s$ asymmetry, then the fit gives
$\kappa = 0.536 \: ^{+\:0.109}_{-\:0.079}$. This was echoed by meson
cloud models, such as Ref.\cite{car00}, which gives $\kappa=0.55$.

The definition of $\kappa$ is
\begin{equation}
\kappa = \frac{\int_{0}^{1}[xs(x) +
x\bar{s}(x)]dx}{\int_{0}^{1}[x\bar{u}(x) + x\bar{d}(x)]dx}.
\end{equation}
Since both experiment \cite{baz95} and our model calculation found
that the strange sea $x$-dependence is similar to that of the
non-strange sea, the $\kappa$ reflects roughly the ratio of strange
quark content relative to the non-strange sea quarks, i.e.,
\begin{equation}
\kappa \approx \frac{2P_{s\bar s}}{P_{u\bar u}+P_{d\bar d}}.
\end{equation}
Then the constraint $\kappa\approx 0.5$, together with $P_{d\bar
d}-P_{u\bar u} =12\%$ and $P_{d\bar d}+P_{u\bar u}+P_{s\bar s} < 1$,
limits the strange quark content in the proton to be in the range of
$P_{s\bar s} = (3\sim 20)\%$. The lower limit $P_{s\bar s} = 3\%$
corresponds to $P_{u\bar u}=0$.

In our model calculation we assume that the $uudd\bar{d}$ system
stays in the configuration of the lowest energy \cite{Helminen}
where the $\bar{d}$ is in its ground state and the $uudd$ subsystem
has mixed orbital symmetry $[31]_{X}$.  This configuration gives the
possibility of 1/4 for $d$ to be in P-wave hence leads naturally to
spin flip.

\section{The single spin asymmetry }

Considering the electroproduction process $\gamma N^{\uparrow}\to
\pi X$ in the five-quark component model of nucleon, the basic
elcetroproduction reaction is $\gamma + q(p) \to \pi + q'(p')$,
where $q'$ represents the final quark which joins in the remaining
part of the nucleon to form the hadronic state X, as illustrated in
Fig. 1.

\begin{figure}[h]
\label {fig1} \centering \mbox{\psfig{figure=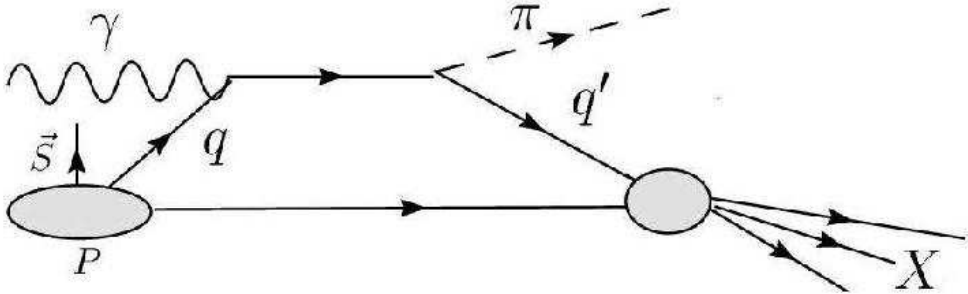,height=0.2
\textheight, clip=}} \caption{Graph for the basic pion
electroproduction reaction of $\gamma q(P)\rightarrow \pi q'(X)$.}
\end{figure}.

The $u$ and $d$ can be taken as the probed quarks in the nucleon,
while the rest partons can be taken as the spectators collectively
in the scattering process. The model assumes that the lifetime of
the five-quark components is much longer than the interaction time
in the scattering process. This makes it possible to describe the
formation of the configuration independently of the measuring
process. The single probed quark can be isolated from the other
quarks in the proton and the information of the strong interaction
between the probed quarks and the spectators in the proton would be
absorbed in the probed quarks wave functions.

The scattering amplitude for the above process can be expressed as
\begin{equation}
A (p,s,k,\lambda; k',s') = \bar{u}(p',s')i
\frac{g_{A}}{f}\gamma_{5}\frac{1}{\slashed{q} - m +
i\epsilon}\slashed{\varepsilon}ie_{q}u(p,s),
\end{equation}
where $k$, $k'$ are the energy-momentum vectors for the photon and
pion, respectively; $p$, $p'$ are the energy-momentum vectors of
initial quark and final quark, respectively; $s$, $s'$ are spins of
them, and $\lambda$ represents the polarization state of photon; $q
= p + k = p' + k'$ and $p'=p+k-k'$. Here, we set the coupling
constant $\frac{g_{A}}{f}$ to be $\frac{1}{f_{\pi}}$. $e_{q}$ is the
electric charge of the incident quark. $u(p,s)$ and
$\bar{u}(p,s)=u^{+}(p,s)\gamma^{0}$ represent the Dirac spinors,
which can be written as the following,
\begin{equation}
\renewcommand{\arraystretch}{0.8}
\bar{u}(p,s) = N\,\left( \begin{array}{cc} 1 \\
\frac{\vec{\sigma}\cdot\vec{p}}{E+m}
\end{array} \right)\chi_{s},\,\,\,u^{+}(p,s) = N\,\chi^{+}_{s}\left( \begin{array}{cc} 1
& \frac{\vec{\sigma}\cdot\vec{p}}{E+m}
\end{array} \right),
\end{equation}
where the normalization factor $N=\sqrt{\frac{E+m}{2m}}$, $\chi_{s}$
the spin vectors of the spin-1/2 quark,
\begin{equation}
~~~~\chi_{1} = \left( \begin{array}{cc} 1 \\
0
\end{array} \right),\,\,\,\,\chi_{2}= \left( \begin{array}{cc} 0\\
1
\end{array} \right).
\end{equation}

For simplicity of concrete calculation, the moving direction of the
photon is taken to be the x direction with two independent photon
polarization vectors $\varepsilon_{\mu}^{1}$ and
$\varepsilon_{\mu}^{2}$ taken as the following,
\begin{equation}
\renewcommand{\arraystretch}{0.8}
 {\varepsilon}_{\mu}^{1} = \left( \begin{array}{cc} 0 \\ 0 \\
                                 1 \\ 0
                                       \end{array} \right), ~~~~~~~
   {\varepsilon}_{\mu}^{2} = \left(
                                  \begin{array}{cc} 0 \\ 0 \\ 0
                                    \\ 1
                                 \end{array} \right).
\end{equation}

Then the basic $\gamma + q(p) \to \pi + q'(p')$ scattering
amplitudes can be calculated as the following for the case of
$\varepsilon_{\mu}=\varepsilon_{\mu}^{1}$:

\begin{eqnarray}
A(\Uparrow\rightarrow\uparrow) &=&
[\frac{2p_{y}p_{z}}{E+m}-\frac{2p_{y}p_{z}-2p^{y}k'_{z}}{E'+m}-
\frac{E_{\gamma}(2p_{y}p_{z}-p_{z}k'_{y}-p_{y}k'_{z})}{(E'+m)(E+m)}]\nonumber\\
&&+ i[\frac{p_{z}k_{x}-k'_{z}k_{x}}{E'+m}
-\frac{p_{z}k_{x}}{E+m}-\frac{E_{\gamma}(p_{x}k'_{z}-p_{z}k'_{x})}{(E'+m)(E+m)}]\nonumber\\
A(\Uparrow\rightarrow\downarrow) &=&
-[\frac{3p_{y}k_{x}-2p_{y}k'_{x}-k_{x}k'_{y}}{E'+m}+
\frac{p_{y}k_{x}-2p_{y}p_{x}}{E+m}\nonumber\\
&&+\frac{E_{\gamma}(2p_{x}p_{y}+p_{y}k_{x}-p_{y}k'_{x}-p_{x}k'_{y})}{(E'+m)(E+m)}]\nonumber\\
&&+i\{\frac{p_{x}k_{x}-2p_{y}^{2}+2p_{y}k'_{y}-k_{x}^{2}-k_{x}k'_{x}}{E'+m}+\frac{p_{x}k_{x}+2p_{y}^{2}}{E+m}\nonumber\\
&&+E_{\gamma}[1+\frac{p_{x}^{2}-p_{y}^{2}+p_{z}^{2}+p_{x}k_{x}-p_{x}k'_{x}+p_{y}k'_{y}-p_{z}k'_{z}}{(E'+m)(E+m)}]\}\nonumber\\
A(\Downarrow\rightarrow\uparrow) &=&
-[\frac{3p_{y}k_{x}-2p_{y}k'_{x}-k_{x}k'_{y}}{E'+m}+
\frac{p_{y}k_{x}-2p_{y}p_{x}}{E+m}\nonumber\\
&&+\frac{E_{\gamma}(2p_{x}p_{y}+p_{y}k_{x}-p_{y}k'_{x}-p_{x}k'_{y})}{(E'+m)(E+m)}]\nonumber\\
&&-i\{\frac{p_{x}k_{x}-2p_{y}^{2}+2p_{y}k'_{y}-k_{x}^{2}-k_{x}k'_{x}}{E'+m}+\frac{p_{x}k_{x}+2p_{y}^{2}}{E+m}\nonumber\\
&&+E_{\gamma}[1+\frac{p_{x}^{2}-p_{y}^{2}+p_{z}^{2}+p_{x}k_{x}-p_{x}k'_{x}+p_{y}k'_{y}-p_{z}k'_{z}}{(E'+m)(E+m)}]\}\nonumber\\
A(\Downarrow\rightarrow\downarrow)
&=&-[\frac{2p_{y}p_{z}}{E+m}-\frac{2p_{y}p_{z}-2p^{y}k'_{z}}{E'+m}-
\frac{E_{\gamma}(2p_{y}p_{z}-p_{z}k'_{y}-p_{y}k'_{z})}{(E'+m)(E+m)}]\nonumber\\
&&+ i[\frac{p_{z}k_{x}-k'_{z}k_{x}}{E'+m}
-\frac{p_{z}k_{x}}{E+m}-\frac{E_{\gamma}(p_{x}k'_{z}-p_{z}k'_{x})}{(E'+m)(E+m)}].
\end{eqnarray}

The label $\Uparrow/\Downarrow$ gives the spin projection $s^{z}_{q}
= \pm \frac{1}{2}$ of the incident quark, and $\uparrow/\downarrow$
corresponding to the spin of final quark $s^{z}_{q'} = \pm
\frac{1}{2}$. $E$ is the energy of incident quark, and $E'$ is that
of the final quark $q'$, where
$E'=E+E_{\gamma}-E_{\pi}=E+E_{\gamma}(1-z)$ with
$z=\frac{E_{\pi}}{E_{\gamma}}$.

The results for the case of
$\varepsilon_{\mu}=\varepsilon_{\mu}^{2}$ can be obtained similarly
as follows.

\begin{eqnarray}
A(\Uparrow\rightarrow\uparrow) &=&
\{\frac{2p_{z}^{2}+p_{x}k_{x}}{E+m}-
\frac{2p_{z}^{2}-p_{x}k_{x}-k_{x}^{2}-2p_{z}k'_{z}+k_{x}k'_{x}}{E'+m}\nonumber\\
&&+E_{\gamma}[\frac{\vec{p}^{2}+p_{x}k_{x}-
p_{x}k'_{x}-p_{y}k'_{y}-p_{z}k'_{z}}{(E'+m)(E+m)}-1]\}\nonumber\\
&&+i[\frac{p_{y}k_{x}}{E+m}-\frac{p_{y}k_{x}-k_{x}k'_{y}}{E'+m}-\frac{E_{\gamma}(p_{y}k'_{x}-p_{x}k'_{y})}{(E'+m)(E+m)}]\nonumber\\
A(\Uparrow\rightarrow\downarrow) &=& [\frac{2p_{x}p_{z}-p_{z}k_{x}}{E+m}-\frac{3p_{z}k_{x}-2p_{z}k'_{x}-k_{x}k'_{z}}{E'+m}
-\frac{E_{\gamma}(p_{z}k_{x}+p_{x}k'_{z}-p_{z}k'_{x})}{(E'+m)(E+m)}]\nonumber\\
&&+i[\frac{2p_{y}p_{z}}{E+m}-\frac{2p_{y}p_{z}-2p_{z}k'_{y}}{E'+m}-\frac{E_{\gamma}(p_{y}k'_{z}-p_{z}k'_{y})}{(E'+m)(E+m)}]\nonumber\\
A(\Downarrow\rightarrow\uparrow) &=&
[\frac{2p_{x}p_{z}-p_{z}k_{x}}{E+m}-\frac{3p_{z}k_{x}-2p_{z}k'_{x}-k_{x}k'_{z}}{E'+m}
-\frac{E_{\gamma}(p_{z}k_{x}+p_{x}k'_{z}-p_{z}k'_{x})}{(E'+m)(E+m)}]\nonumber\\
&&-i[\frac{2p_{y}p_{z}}{E+m}-\frac{2p_{y}p_{z}-2p_{z}k'_{y}}{E'+m}-\frac{E_{\gamma}(p_{y}k'_{z}-p_{z}k'_{y})}{(E'+m)(E+m)}]\nonumber\\
A(\Downarrow\rightarrow\downarrow)
&=&-\{\frac{2p_{z}^{2}+p_{x}k_{x}}{E+m}-
\frac{2p_{z}^{2}-p_{x}k_{x}-k_{x}^{2}-2p_{z}k'_{z}+k_{x}k'_{x}}{E'+m}\nonumber\\
&&+E_{\gamma}[\frac{\vec{p}^{2}+p_{x}k_{x}-
p_{x}k'_{x}-p_{y}k'_{y}-p_{z}k'_{z}}{(E'+m)(E+m)}-1]\}\nonumber\\
&&+i[\frac{p_{y}k_{x}}{E+m}-\frac{p_{y}k_{x}-k_{x}k'_{y}}{E'+m}-\frac{E_{\gamma}(p_{y}k'_{x}-p_{x}k'_{y})}{(E'+m)(E+m)}].
\end{eqnarray}

We can obtain the cross section for the scattering of photon and
quark from the above amplitudes as follows,
\begin{equation}
d\sigma_{\gamma q}(p,s,k; k')=
\sum_{\lambda}\sum_{s'}\frac{(2\pi)^{4}m_{q}}{2(p\cdot
k)}|A(p,s,k,\lambda;k',s')|^{2}\frac{d^{3}k'}{(2\pi)^{3}2E_{\pi}}\frac{m_{q}}{(2\pi)^{3}(E_{q}+E_{\gamma}-E_{\pi})}.
\end{equation}
Here $d\sigma_{\gamma q}$ describes the cross section of the
scattering between photon and the quark of the momentum $p$. In the
proton, the momentum of quark should have a density distribution
$\rho_q(p)$. Then the scattering cross section of photon and proton
can be obtained by summing over quarks inside the proton as
\begin{equation}
d\sigma_{\gamma p}(x,k;k')=\sum_{q}\int d^{4}p_{q}
\rho_q(p_q)d\sigma_{\gamma q}(p_q,s,k;k')\delta(p_{q}^{+}-xP^{+})
\label{eq1}
\end{equation}
where $p^+_q$ and $P^+$ are the light-cone momenta of the struck
quark and proton, respectively. By integrating over the $k'_{x}$,
$E_{\pi}$ and $|k'_{T}|$, the differential cross section vs the
angle $\phi$ of the produced pion can be obtained as
\begin{equation}
d\sigma_{\gamma p}(x;\phi)=\int
dE_{\pi}dk'_{x}d|k'_{T}|d\sigma_{\gamma p}(x,k;k').\label{eq2}
\end{equation}

The single spin asymmetry (SSA) observable for the $\gamma p\to\pi
X$ reaction is defined as
\begin{equation}
A_{UT}(x,\phi) = \frac{d \sigma ^{\uparrow}(x;\phi) - d \sigma
^{\downarrow}(x;\phi)}{d \sigma ^{\uparrow}(x;\phi) + d \sigma
^{\downarrow}(x;\phi)}.
\end{equation}
where $d \sigma ^{\uparrow}(x;\phi)$ ($d \sigma
^{\downarrow}(x;\phi)$) represents the differential cross section of
the scattering between photon and the proton polarized upwards
$J_{p}^{z}=+\frac{1}{2}$ (downwards $J_{p}^{z}=-\frac{1}{2}$).

In the constituent quark model, the quark momentum density
distribution function $\rho_q(p)$ is proportional to the square of
the quark wave function. The simple harmonic oscillator wave
functions are commonly used with radial parts for the S-state and
P-state as
\begin{eqnarray}
\varphi^{S}(p)&=&{1\over (\alpha^2\pi)^{3/4}}
\exp(-\frac{p^{2}}{2\alpha^{2}}),\\
\varphi^{P}(p)&=&{p\over \alpha}\,\varphi^{S}(p)\, ,
\end{eqnarray}
respectively. Here $\alpha^{2} = m_{q}\omega$ with $\omega$ the
harmonic oscillator parameter. While it is isotropic for the
S-state, the angular dependent part is given by $Y_{1l_z}$ for the
P-state with $l_z=1,0,-1$ for the quark polarization projection
along the proton polarization direction ({\sl i.e.}, $z$ axis).

In the conventional quark model with 3 constituent quarks, all 3
quarks are in the ground S-states. With above formulae, the $\gamma
p\to\pi X$ differential cross section does not depend on the
polarization of the proton and hence does not result in any
asymmetry. On the other hand, in the five-quark component
model~\cite{zou05,an206,ris06,an06}, there are at least 12\%
$uudd\bar d$ component in the proton with the anti-quark in the
orbital ground state and the four quarks in the mixed orbital
$[31]_X$ symmetry, i.e., one in P-wave and three in S-wave, together
with flavor-spin $[4]_{FS}[22]_F[22]_S$ symmetry.  When the struck
quark is in P-state, there will be orbital-spin coupling which leads
to quark spin-flip. The orbital-spin coupling will give nonzero
single spin asymmetry. For the $[31]_X[4]_{FS}[22]_F[22]_S$
configuration of the $uudd\bar d$ component, when
$J_{p}^{z}=+\frac{1}{2}$ there is a probability of 2/3 with $l_z=+1$
and 1/3 with $l_z=0$; when $J_{p}^{z}=-\frac{1}{2}$ there is a
probability of 2/3 with $l_z=-1$ and 1/3 with $l_z=0$. From these
different probabilities for various $l_z$, with formulae above one
can get different $d \sigma ^{\uparrow}(x;\phi)$ and $d \sigma
^{\downarrow}(x;\phi)$.

For the $\pi^+$ production, the struck quark should be either $u$
quark or $\bar d$ quark. In the five-quark component model with a
portion of $P_{5q}$ for the 5-quark component and $P_{3q}$ for the
3-quark component, the SSA for the $\pi^+$ production should be
\begin{eqnarray}
A^{\pi^+}_{UT}(x;\phi) &=& \frac{P_{5q}[d \sigma_{u(5q)}
^{\uparrow}(x;\phi) - d \sigma
_{u(5q)}^{\downarrow}(x;\phi)]}{P_{5q}[d \sigma_{u,\bar{d}(5q)}
^{\uparrow}(x;\phi)+d \sigma_{u,\bar{d}(5q)}
^{\downarrow}(x;\phi)]+P_{3q}[d \sigma _{u(3q)}^{\uparrow}(x;\phi) +
d \sigma_{u(3q)}
^{\downarrow}(x;\phi)]}\nonumber\\
&=& \frac{d \sigma_{u(5q)} ^{\uparrow}(x;\phi) - d \sigma
_{u(5q)}^{\downarrow}(x;\phi)}{d \sigma_{u,\bar{d}(5q)}
^{\uparrow}(x;\phi)+d \sigma_{u,\bar{d}(5q)}
^{\downarrow}(x;\phi)+2P_{3q}/P_{5q}~d \sigma _{u(3q)}(x;\phi)}
\end{eqnarray}
where the $d\sigma_{u(3q)}$ and $d \sigma_{u,\bar{d}(5q)}$ represent
the cross sections for the scattering processes of photon and $u$
quark in the three-quark proton and $u$, $\bar{d}$ in the 5-quark
proton, respectively. Since the scattering on the 3-quark proton
does not depend on the proton polarization, we have $d \sigma
_{u(3q)}^{\uparrow}(x;\phi) = d \sigma_{u(3q)}
^{\downarrow}(x;\phi)\equiv d \sigma _{u(3q)}(x;\phi)$.

For the high energy $\gamma p\to\pi^+ X$ experiment at
HERMES~\cite{air05}, the $X$ is usually a multi-hadron state. Since
a 5-quark state is easier to transit into a multi-hadron state than
a 3-quark state, the $d \sigma _{u(3q)}$ is expected to be somewhat
smaller than $d \sigma _{u,\bar{d}(5q)}$. In the minimum 5-quark
component model, there is only 12\% $uudd\bar d$ component without
any $u\bar u$ and $s\bar s$ components. We calculate the SSA for the
HERMES $\gamma p\to\pi^+X$ experiment~\cite{air05} vs
$x=\frac{E-p_x}{M_p}$ as shown in Fig.2 by assuming $d \sigma
_{u(3q)}=0$ (solid line) and $d \sigma _{u(3q)}=d \sigma
_{u,\bar{d}(5q)}$ (dotted line), respectively. The range defined by
two curves covers the experimental data well.

\begin{figure}[h]
\label {fig2} \centering
\mbox{\psfig{figure=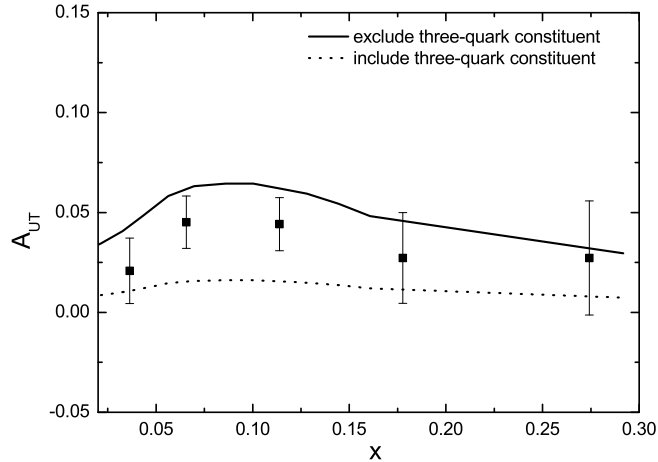,height=0.3 \textheight, clip=}}
\caption{The calculated single-spin asymmetry vs $x$ for
$\gamma^*p\to\pi^+X$ with $90^o\ge\phi\ge 0^o$ by assuming
$d\sigma_{u(3q)}=0$ (solid line) and $d\sigma_{u(3q)}=d
\sigma_{u,\bar d(5q)}$ (dotted line), compared with HERMES
data~\cite{air05}. }
\end{figure}

However, the minimum 5-quark component model gives a nearly same
prediction for the $\pi^-$ production while the HERMES
experiment~\cite{air05} found a different SSA for the $\pi^-$
production from the $\pi^+$ production as shown in Fig.3. The reason
is that there are equal number of $d$ and $u$ quarks for the
$uudd\bar d$ component. So we need to go beyond the minimum 5-quark
component model. We should include the $uuud\bar{u}$ component with
the $uuud$ configuration of the $[4]_{FS}[31]_{F}[31]_{S}$
flavor-spin symmetry, which is likely to have the lowest energy for
the 5-quark component with $u\bar u$~\cite{Helminen}. Such component
has the $u/d$ quark ratio to be 3 and results in larger SSA for
$\pi^+$ production than for $\pi^-$ production. In addition,
according to Eq.(3), there should also be some $uuds\bar s$
component with $P_{s\bar s}\approx (P_{u\bar u}+P_{d\bar d})/4$.
This configuration has the $u/d$ quark ratio to be 2 and should also
result in larger SSA for $\pi^+$ production than for $\pi^-$
production. For the $\pi^-$ production, the SSA should be
\begin{equation}
A^{\pi^-}_{UT}(x;\phi) = \frac{d \sigma_{d(5q)} ^{\uparrow}(x;\phi)
- d \sigma _{d(5q)}^{\downarrow}(x;\phi)}{d \sigma_{d,\bar{u}(5q)}
^{\uparrow}(x;\phi)+d \sigma_{d,\bar{u}(5q)}
^{\downarrow}(x;\phi)+2P_{3q}/P_{5q}~d \sigma _{d(3q)}(x;\phi)}
\end{equation}
In Fig.3, we show the results with 5\% $uuud\bar{u}$, 17\%
$uudd\bar{d}$ and 5.5\% $uuds\bar s$ components, which satisfy
$P_{d\bar d}-P_{u\bar u}=12\%$ and $P_{s\bar s}=(P_{u\bar
u}+P_{d\bar d})/4$, under the assumption of $d\sigma_{(3q)}=d
\sigma_{(5q)}$ for both $\pi^+$ and $\pi^-$ production. Compared
with HERMES data~\cite{air05}, the $\pi^+$ production is very well
reproduced; for the $\pi^-$ production, it is also mostly within the
error bars although it seems systematically larger than the central
values of the data. Reducing the percentage of the $uuud\bar{u}$
component will lower down SSA for both $\pi^+$ and $\pi^-$
production to be closer to the dotted line corresponding to the
results without including $uuud\bar{u}$ and $uuds\bar s$ components.

\begin{figure}[h]
\label {fig3} \centering \mbox{\psfig{figure=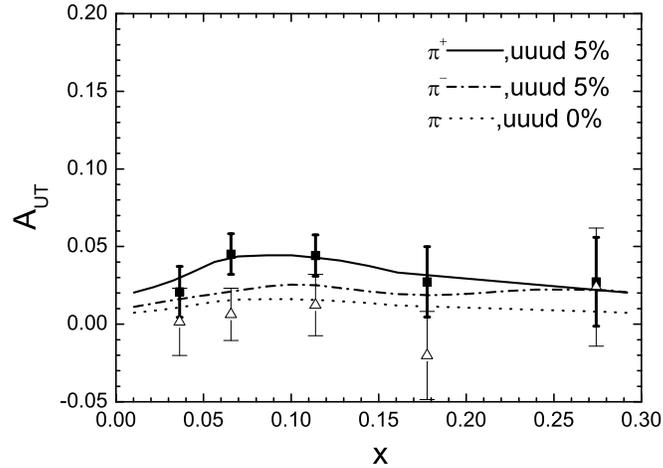,height=0.3
\textheight, clip=}} \caption{The calculated single-spin asymmetry
vs $x$ by assuming $d\sigma_{(3q)}=d \sigma_{(5q)}$ in the 5-quark
component model with 5\% $uuud\bar{u}$, for the $\gamma^*p\to\pi X$
reaction with solid line for $\pi^+$ and dot-dashed line for
$\pi^-$; compared with HERMES data~\cite{air05} (square for $\pi^+$
and triangle for $\pi^-$) and the result without including the
$uuud\bar{u}$ and $uuds\bar s$ components (dotted line).}
\end{figure}

\section{Summary and discussion}

We have calculated the single-spin asymmetry for the pion
leptoproduction process $\gamma^* p^{\uparrow}\to \pi X$ in the
five-quark component model. The orbital-spin coupling, which is the
natural consequence of the five-quark configurations, can be viewed
as the source of the asymmetry. The well established light flavor
sea quark asymmetry in the proton, $\bar d-\bar u
=0.12$~\cite{Garvey}, demands $P_{d\bar d}-P_{u\bar u}=0.12$ and
sets the lower limit for the percentage of $uudd\bar d$ component to
be 12\%. To reproduce the empirical evidence that the SSA for
$\pi^-$ production is smaller than for $\pi^+$ production, non-zero
$P_{u\bar u}$ and/or $P_{s\bar s}$ are needed. With 5\% $uudu\bar
u$, 17\% $uudd\bar{d}$ and 5.5\% $uuds\bar s$ components, which
satisfy the constraints of $P_{d\bar d}-P_{u\bar u}=12\%$ and
$P_{s\bar s}=(P_{u\bar u}+P_{d\bar d})/4$, we can obtain SSA
consistent with recent HERMES data~\cite{air05} for both $\pi^+$ and
$\pi^-$ production, although the prediction of the SSA for $\pi^-$
production seems larger than the central values of the data
systematically. More precise data will be helpful to clarify the
consistence.

The five-quark component model has been applied successfully to
study many properties of proton. A complete
analysis~\cite{zou05,ris06,an06} of the relation between the
strangeness observables and the possible configurations of the
$uuds\bar{s}$ component of the proton concludes that, for a negative
$\Delta_{s}$, positive $\mu_{s}$ and $r_{s}$, the $\bar{s}$ is in
the ground state and the $uuds$ system in P-states. Based on this
configuration, the $s$-$\bar{s}$ asymmetry has been obtained and it
can account for 10-20\% of NuTeV anomaly~\cite{wei08}. In addition,
the light flavor asymmetry of sea quarks can also be obtained by the
five-quark components with percentage determined by the principle of
detailed balance~\cite{zha01}. It gives naturally the quark orbital
angular momentum which is crucial for solving the famous proton spin
crisis and single spin asymmetry. In conclusion, the inclusion of
$20\%\sim 30\%$ five-quark components in the proton helps to
understand many properties of the proton.

\bigskip
\noindent {\bf Acknowledgements:} We thank C.S.An for helpful
discussions. This work is partly supported by the National Natural
Science Foundation of China under grants Nos. 10435080, 10521003 and
by the Chinese Academy of Sciences under project No. KJCX3-SYW-N2.

\end{document}